\documentclass[aps,prb,twocolumn,superscriptaddress]{revtex4-2}

\usepackage{color}
\usepackage{graphicx}
\usepackage{dcolumn}
\usepackage{bm}
\usepackage{float}
\usepackage[mathlines]{lineno}
\usepackage{tabularx}
\usepackage{hyperref}
\usepackage{footnote}
\usepackage{amsmath}
\usepackage{txfonts}

\begin{document}

\title{Superconductivity at ambient pressure in hole-doped LuH$_3$}

\author{Zhenfeng Ouyang}\affiliation{Department of Physics and Key Laboratory of Quantum State Construction and Manipulation (Ministry of Education), Renmin University of China, Beijing 100872, China}
\author{Miao Gao}\email{gaomiao@nbu.edu.cn}\affiliation{Department of Physics, School of Physical Science and Technology, Ningbo University, Zhejiang 315211, China}\affiliation{School of Physics, Zhejiang University, Hangzhou 310058, China}
\author{Zhong-Yi Lu}\email{zlu@ruc.edu.cn}\affiliation{Department of Physics and Key Laboratory of Quantum State Construction and Manipulation (Ministry of Education), Renmin University of China, Beijing 100872, China}

\date{\today}

\begin{abstract}
Very recently, a report on possible room-temperature superconductivity in N-doped lutetium hydrides near 1 GPa pressure has drawn lots of attentions.
To date, the superconductivity is not confirmed under relatively low pressure in subsequent studies.
Based on the density functional theory first-principles calculations, we extensively investigate the influence of charge doping on the stability and superconductivity of LuH$_3$ at ambient pressure.
Although electron doping can not stabilize LuH$_3$,
we find that the room-pressure stability of LuH$_3$ can be achieved by doping holes with concentrations in between 0.15-0.30 holes/cell.
Moreover, our calculations reveal a positive dependence of superconducting transition temperature on the number of doped holes, with the highest value close to 54 K. These findings suggest that realizing superconductivity in hole-doped LuH$_3$ at ambient pressure is not impossible, although the transition temperature is still far away from the room temperature.
\end{abstract}


\maketitle

\section{Introduction}

Finding a superconductor that works at ambient conditions is the ultimate goal of superconductivity research.
Guiding by the Bardeen-Cooper-Schrieffer (BCS) theory \cite{Bardeen-PR108},
conventional superconductor composed of light elements may possess a high transition temperature ($T_c$), owing to its high Debye temperature, which reflects the energy scale of electron pairing.
As the lightest element, high-$T_c$ even room-temperature superconductivity has been anticipated in extremely compressed solid metallic hydrogen for a long period.
Facing the difficulty of metallizing solid hydrogen, N. Ashcroft proposed that high-$T_c$ superconductivity may exist in hydrogen-rich materials at relatively low pressure, based on the effect of chemical precompression \cite{Ashcroft-PRL92}. Recent years, taking the results of crystal structure prediction as references, near-room-temperature superconductivity in hydrogen-rich compounds has been observed in experment above megabar pressure, such as H$_3$S \cite{Duan-SR4,Drozdov-nature525}, LaH$_{10}$ \cite{Liu-PNAS114,Drozdov-nature569}, and YH$_9$ \cite{Liu-PNAS114,Peng-PRL119,Snider-PRL126}. Ternary hydrides are regarded as potential high-$T_c$ superconductors at relatively low pressure, for example LaBH$_8$ \cite{Cataldo-PRB104,Zhang-PRL128,Liang-PRB104}, LaBeH$_8$ \cite{Zhang-PRL128}, KB$_2$H$_8$ \cite{Gao-PRB104}, H$_6$SCl \cite{Hai-PRB105}, CsBH$_5$ \cite{Gao-PRB107}, SrSiH$_8$ \cite{Lucrezi-npj8}, and (La,Ce)H$_{9-10}$ \cite{Chen-NC14}, but the reduction of critical pressure always leads to depressed $T_c$.

Very recently, Dias \emph{et al.} claimed the discovery of superconductivity in N-doped lutetium hydrides at 1 GPa, with $T_c$ being 294 K. The main phase that accounts for superconductivity was attributed to $Fm\bar{3}m$ LuH$_{3-\delta}$N${_\epsilon}$, in which a color change from blue to pink and red occurs below 4 GPa \cite{Dasenbrock-Nature615}.
Although LuH$_3$ at 122 GPa and Lu$_4$H$_{23}$ at 218 GPa were reported to be superconductive \cite{Shao-IC60,Li-SCPMA66}, the signature of room-temperature superconductivity in Lu-N-H close to ambient pressure has not yet been extensively confirmed. From 1 GPa to 6 GPa, the metallic behavior of $Fm\bar{3}m$ LuH$_{2\pm x}$N$_y$ shows progressive optimization, but no superconductivity down to 10 K \cite{Ming-nature}.
The pressure-induced color changes of LuH$_{2\pm x}$N$_y$ was observed under much higher pressure from 10 to 30 GPa, without finding the trace of superconductivity \cite{Zhang-SCPMA66,Xing-arXiv}.
It is noteworthy that the color variation of LuH$_2$ behaves similarly to N-doped lutetium hydride \cite{Shan-CPL40}. However, the evidence of superconducting transition in LuH$_2$ was not found down to 1.5 K, with pressure up to 7 GPa. Cai \emph{et al.} synthesized the Lu-N-H samples using the same experimental method reported by Dias \emph{et al.} but superconducting transition was not observed in the measurements of resistance and magnetic susceptibility around the near-ambient pressure \cite{Cai-arXiv}.
Hemley's group emphasized the appearance of superconductivity may be strongly dependent on synthesis conditions and claimed the discovery of superconductivity in the samples synthesized by Dias \emph{et al} \cite{Salke-arXiv}.
Based on the density functional theory (DFT) calculations, X-ray diffraction and optical properties of lutetium hydrides were simulated. It was suggested that the CaF$_2$-type LuH$_2$ holds a dominant position, with a minor phase of NaCl-type LuH \cite{Liu-arXiv}. DFT calculations revealed that there are no stable Lu-N-H ternary structures \cite{Xie-CPL40,Huo-arXiv,Sun-arXiv,Hilleke-arXiv,Gubler-arXiv}. The dynamical stability and superconductivity of N-doped LuH$_3$ were studied through both supercell method and virtual crystal approximation (VCA). It was suggested that $R3m$ Lu$_2$H$_5$N can become superconductive under 50 GPa \cite{Huo-arXiv}. According to the VCA results, the critical pressures to stabilize N-doped LuH$_3$ are equal to 25 GPa and 70 GPa for N concentration being 1\% and 2\%, respectively. The highest $T_c$ is 22 K in VCA, which can be achieved in 1\% N-doped LuH$_3$ under 30 GPa. Similarly, a CaF$_2$-type LuNH phase was proposed to be superconducting under 1 GPa, with $T_c$ reaching 17 K \cite{Hilleke-arXiv}.

\begin{figure}[tbh]
\centering
\includegraphics[width=8.6cm]{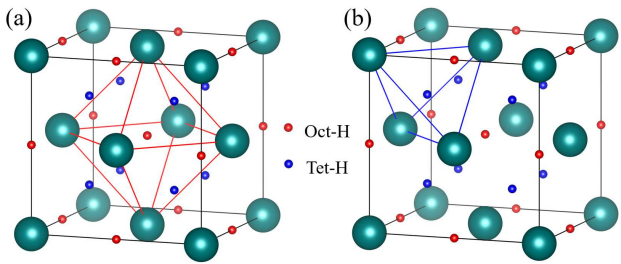}
\caption{Crystal structure for LuH$_3$. Lu atoms are in dark cyan. The red and blue balls represent two unequivalent H atoms locating at the octahedral and tetrahedral interstitial sites, respectively. The solid red line in (a) and blue line in (b) denotes the octahedron and tetrahedron formed by Lu atoms. }
\label{fig:Stru}
\end{figure}

It is known that $Fm\bar{3}m$ LuH$_3$ is unstable at ambient pressure. To date, near-room-pressure stability of N-doped LuH$_3$ has not yet been realized either experimentally or theoretically.
Thus, it is crucial to examine whether LuH$_3$ can be a stable compound, even a superconductor, at ambient pressure, via special doping condition.
In this work, based on DFT calculations, we systematically investigate the influence of doping effect on the dynamical stability and phonon-mediated superconductivity in LuH$_3$, with doping level ranging from 0.50 electrons/cell to 0.50 holes/cell. As revealed by our calculations, the maximal value of the imaginary phonon frequency increases with the concentration of electron doping, reflecting its negative effect in stabilizing LuH$_3$. Conversely, the dynamical stability can be markedly improved by introducing holes. In particular, the imaginary phonon frequencies disappear, when the hole doping concentration locating between 0.15 to 0.30 holes/cell. At these doping level, the electronic structure and phonons are calculated. In order to accurately ascertain the electron-phonon coupling (EPC), we adopt the Wannier interpolation technique. It is found that phonons associated with Lu and H atoms at the octahedral centers have strong coupling with electrons.
By solving the isotropic Eliashberg equations, the highest $T_c$ of 53.7 K can be achieved in 0.30 holes/cell-doped LuH$_3$.
The enhanced EPC matrix elements of Lu phonons and the softened phonons of hydrogen atoms locating at the octahedral centers can account for the positive relationship between $T_c$ and the number of doped holes.

\section{Method}
The plane-wave method, {\small QUANTUM-ESPRESSO} \cite{Giannozzi-JPCM21} package was adopted in our DFT calculations.
The Perdew-Burke-Ernzerhof formula was selected as the exchange and correlation functional \cite{Perdew-PRL77}. The electron-ion interactions were treated by Rappe-Rabe-Kaxiras-Joannopoulos ultrasoft pseudopotentials.
The $4f$ electrons were not included in valence configuration, since the full-filled $4f$ orbitals are buried deeply below the Fermi level and can hardly
affect the EPC.
After convergence test, the kinetic energy cutoff and the charge density cutoff were chosen to be 80 Ry and 320 Ry, respectively.
When calculating the self-consistent charge densities, the Brillouin zone was sampled by a {\bf k} mesh of 18$\times$18$\times$18 points with a Methfessel-Paxton smearing \cite{Methfessel-PRB40} of 0.02 Ry.
The electron/hole doping was simulated by adding/removing a certain amount of electrons from the
system. Meanwhile, a compensating background of uniform positive/negative charges was introduced to avoid numerical divergence.
The dynamical matrices and the perturbation potential were calculated on a mesh of 6$\times$6$\times$6 points, within the framework of density-functional perturbation theory \cite{Baroni-RMP73}.

The maximally localized Wannier functions \cite{Pizzi-JPCM32} (MLWFs) were constructed on a 6$\times$6$\times$6 grid of the Brillouin zone.
Three $s$ orbitals of H atoms were used as an initial guess. The convergent EPC constant $\lambda$ was extensively carried out through fine electron (72$\times$72$\times$72) and phonon (24$\times$24$\times$24) grids with Electron-Phonon Wannier (EPW) code \cite{Ponce-CPC209}. Specifically, the dirac $\delta$ functions for electrons and phonons were smeared out by a Gaussian function with the widths of 90 meV and 0.5 meV, respectively. The fine electron grid of 72 $\times$ 72 $\times$ 72 points was employed when solving the isotropic Eliashberg equations \cite{Ponce-CPC209, Choi-PC385, Margine-PRB87}. The sum over the Matsubara frequencies was truncated with $\omega_{c}$ = 1.6 eV, about ten times that of the highest phonon frequency.

\section{Results and discussion}

\begin{figure}[tbh]
\centering
\includegraphics[width=8.6cm]{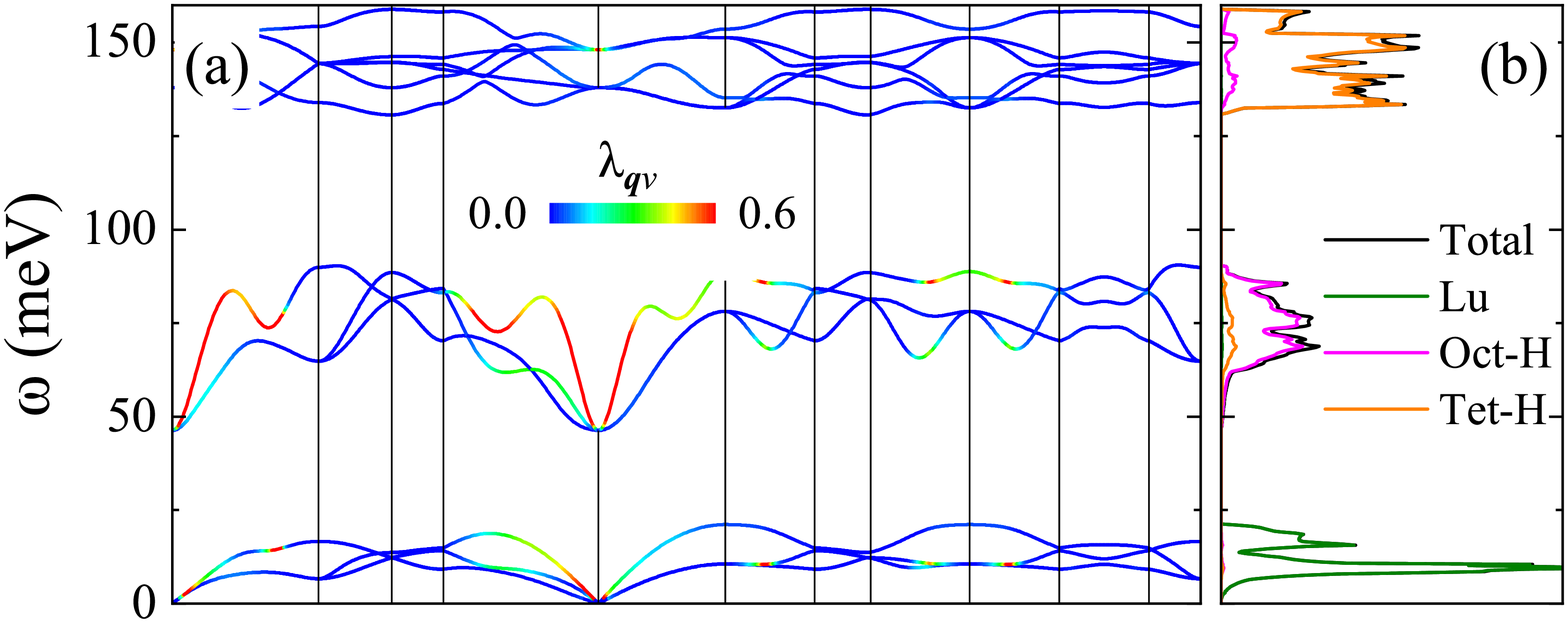}
\includegraphics[width=8.6cm]{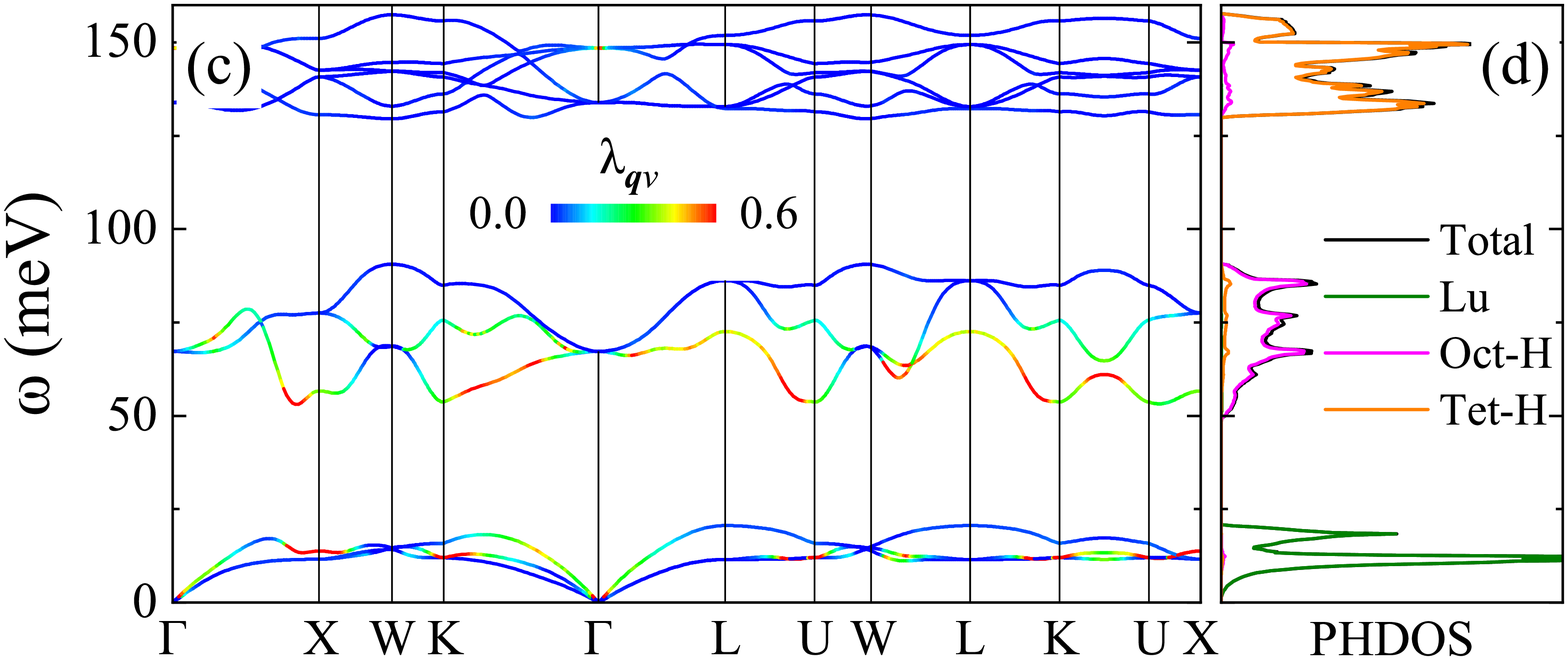}
\caption{Lattice dynamics of hole-doped LuH$_3$. Phonon spectrum with a color representation of $\lambda_{{\bf q}v}$ for (a) 0.15 holes/cell and (c) 0.30 holes/cell. Projected phonon DOS generated by quasiharmonic approximation for (b) 0.15 holes/cell and (d) 0.30 holes/cell.  }
\label{fig:phonon}
\end{figure}

The crystal structure of $Fm\bar{3}m$ LuH$_3$ is shown in Fig. \ref{fig:Stru}.
There are two unequivalent H atoms, which occupy $4b$ (0.500 0.000 0.000) and $8c$ (0.250 0.250 0.250) Wyckoff positions, locating at the octahedral and tetrahedral interstitial
sites, respectively.
For clarity, these two atoms are labeled as Oct-H and Tet-H.
At ambient pressure, the lattice constants determined for undoped LuH$_3$ is 4.9739 \AA, in good agreement with the previous theoretical values \cite{Liu-arXiv}.
The largest imaginary phonon frequency of LuH$_3$ is about -50.11 meV, in excellent agreement with previous results \cite{Dasenbrock-Nature615, Liu-arXiv}, revealing its instability at ambient pressure.
We firstly calculate the phonon spectra under various concentrations of electron doping, from 0.05 electrons/cell to 0.50 electrons/cell, with doping step being 0.05 electrons/cell.
Accompanied by the increase of electron doping, the maximal imaginary phonon frequencies are equal to -65.58, -70.93, -72.83, -73.46, -76.75, -81.71, -86.33, -90.79, and -99.11 meV, respectively.
This suggests that the ambient-pressure stability of LuH$_3$ can not be realized through doping electrons.
Therefore, it is quite interesting to investigate whether hole doping can significantly enhance the dynamical stability.

Likewise, the number of doped holes is ranging from 0.05 holes/cell to 0.50 holes/cell.
By examining the phonon spectrum,
we find that the imaginary modes disappear with doping level in between 0.15 holes/cell and 0.30 holes/cell [Fig.~\ref{fig:phonon}].
The phonon spectrum is divided into three parts by two frequency gaps [Fig.~\ref{fig:phonon}(b)]. According to the projected phonon DOS, Lu atoms mainly participate in the vibrations below 25 meV.
Three optical modes below 100 meV are associated with Oct-H atoms. Tet-H atoms mainly contribute to the remanent six optical modes above 100 meV.
The Oct-H related phonons are sensitive to hole doping. Although the optical modes at the $\Gamma$ point are slightly hardened, most Oct-H phonons show obvious softening.
Visible softening can also be seen for Lu phonons, especially near the $X$ and $K$ points [Fig.~\ref{fig:phonon}(c)].
In comparison, the response of Tet-H phonons to the hole concentration is negligible.

\begin{figure}[tbh]
\centering
\includegraphics[width=8.6cm]{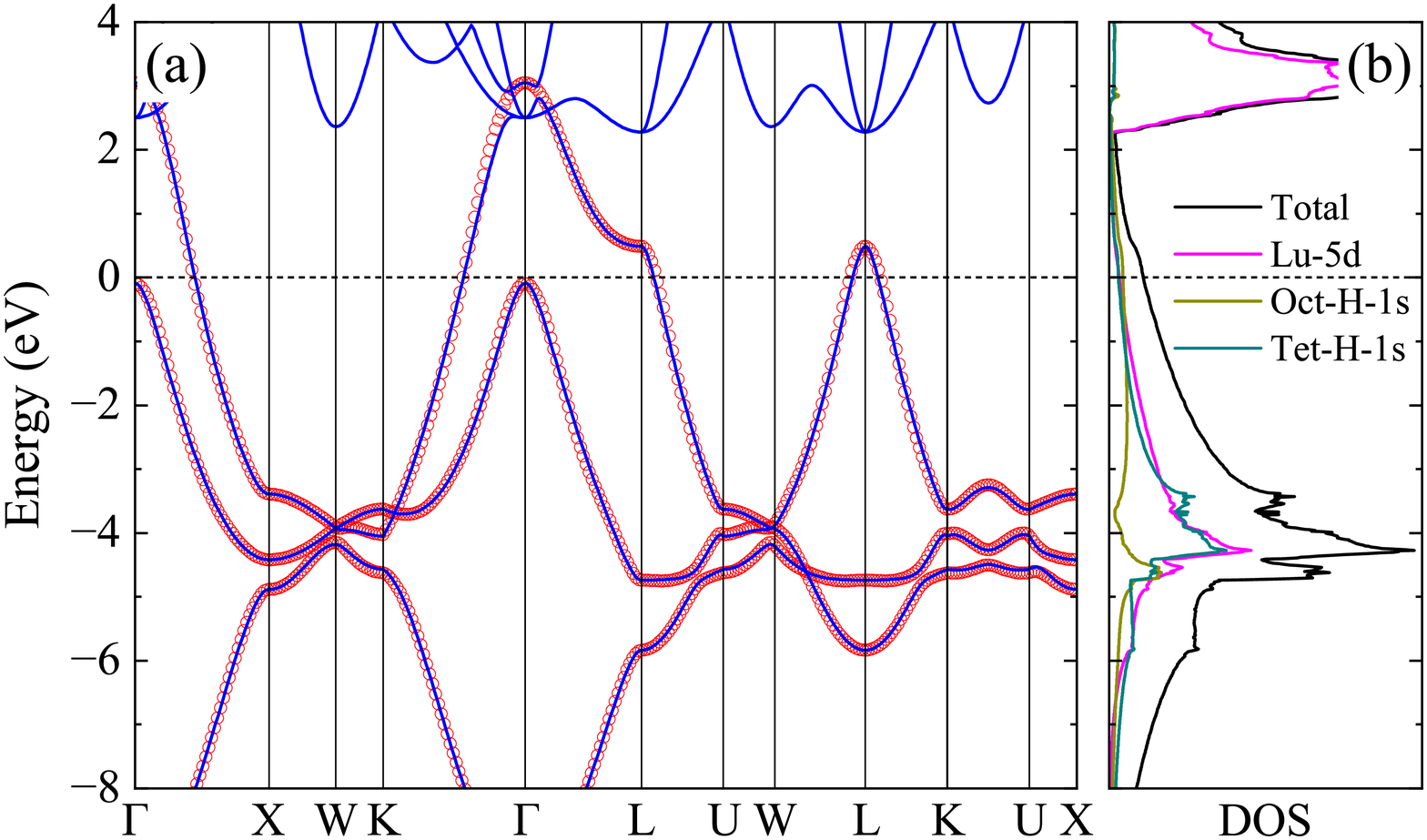}
\includegraphics[width=8.6cm]{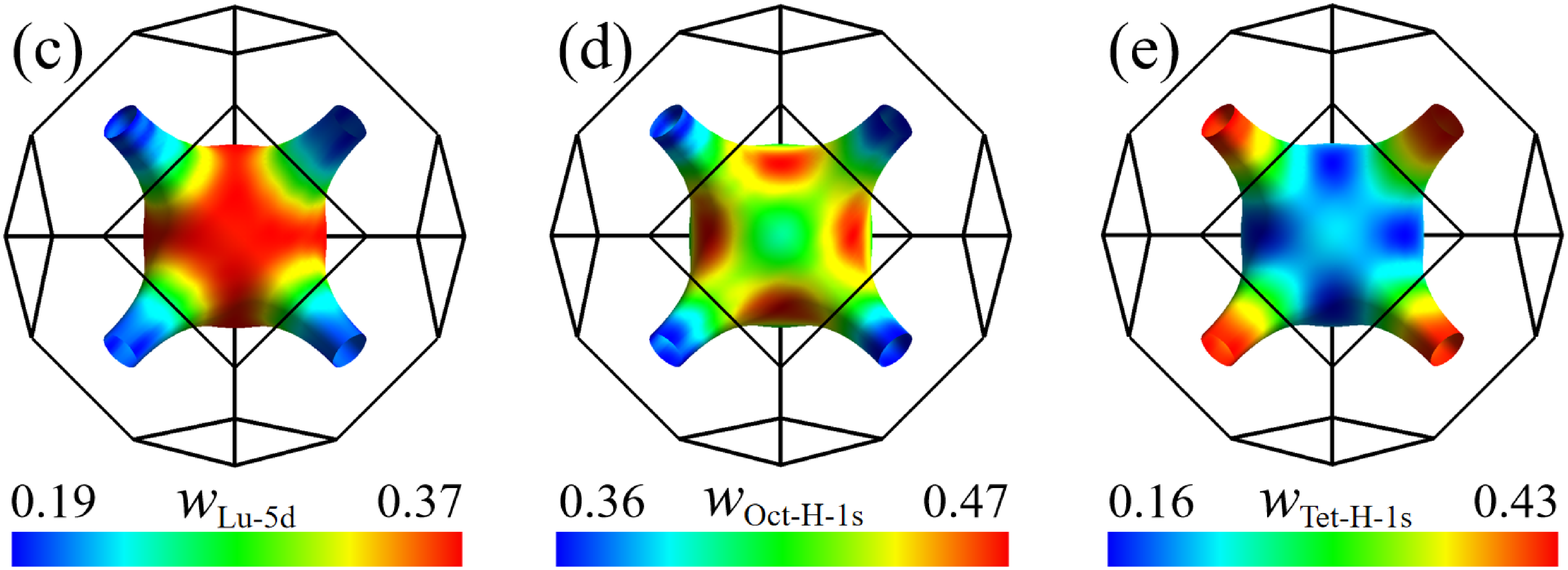}
\caption{Electronic structure of 0.30 holes/cell doped LuH$_3$. (a) Band structure. The blue lines and red circles denote the energy bands obtained by first-principles calculation and Wannier interpolation, respectively. The Fermi level is set to zero. (b) Total and orbital-resolved DOS. (c)-(e) Distributions of different orbitals on the Fermi surfaces.}
\label{fig:band}
\end{figure}

Since room-pressure stability of LuH$_3$ has been acquired via hole doping, we thus investigate its possible superconductivity within the phonon-mediated mechanism.
Figure \ref{fig:band} shows the electronic structure of LuH$_3$ under 0.30 holes/cell doping.
Only one band is across the Fermi level. Interestingly, the Fermi level touches the maximum of another valance band at the $\Gamma$ point [Fig.~\ref{fig:band}(a)]. Further raising the doping level will result in partial occupation of such band. This may be closely related to the lattice instability at higher doping.
To confirm the validity of subsequent Wannier interpolation, the consistency between DFT bands and the MLWFs interpolated ones is carefully checked.
The main peak of density of states (DOS) locates around -4.0 eV [Fig.~\ref{fig:band}(b)]. For the filled states, Tet-H-$1s$ and Lu-$5d$ orbitals possess almost equal contribution, reflecting strong chemical bonding between Tet-H and Lu atoms, since Tet-H is the nearest neighbor of Lu. This can also account for the domination of Tet-H in the high-frequency optical phonon modes [Fig.~\ref{fig:phonon}(b) and Fig.~\ref{fig:phonon}(d)].
For the DOS at the Fermi level, i.e. $N(0)$, Oct-H-$1s$ slightly overwhelms these two orbitals.
According to the profile of total DOS, $N(0)$ shows a monotonic increase with the number of doped holes. For example, $N(0)$ equals 0.078, 0.116, 0.123, 0.128 states/spin/eV/cell, for doping level being
0.15, 0.20, 0.25, and 0.30 holes/cell, respectively.
Other orbitals of Lu, such as $4p$, $5s$, $5p$, and $6s$ are not given in Fig.~\ref{fig:band}(b), since they have insignificant contribution to the DOS from -4.0 eV to 4.0 eV.
The Fermi surfaces have a cubic shape, with eight antennas directing the $L$ point, the center of hexagonal boundary face of Brillouin zone.
The cubic part, the saddle points, and the antennas are dominated by Lu-$5d$, Oct-H-$1s$, and Tet-H-$1s$ orbitals.

\begin{figure}[tbh]
\centering
\includegraphics[width=8.6cm]{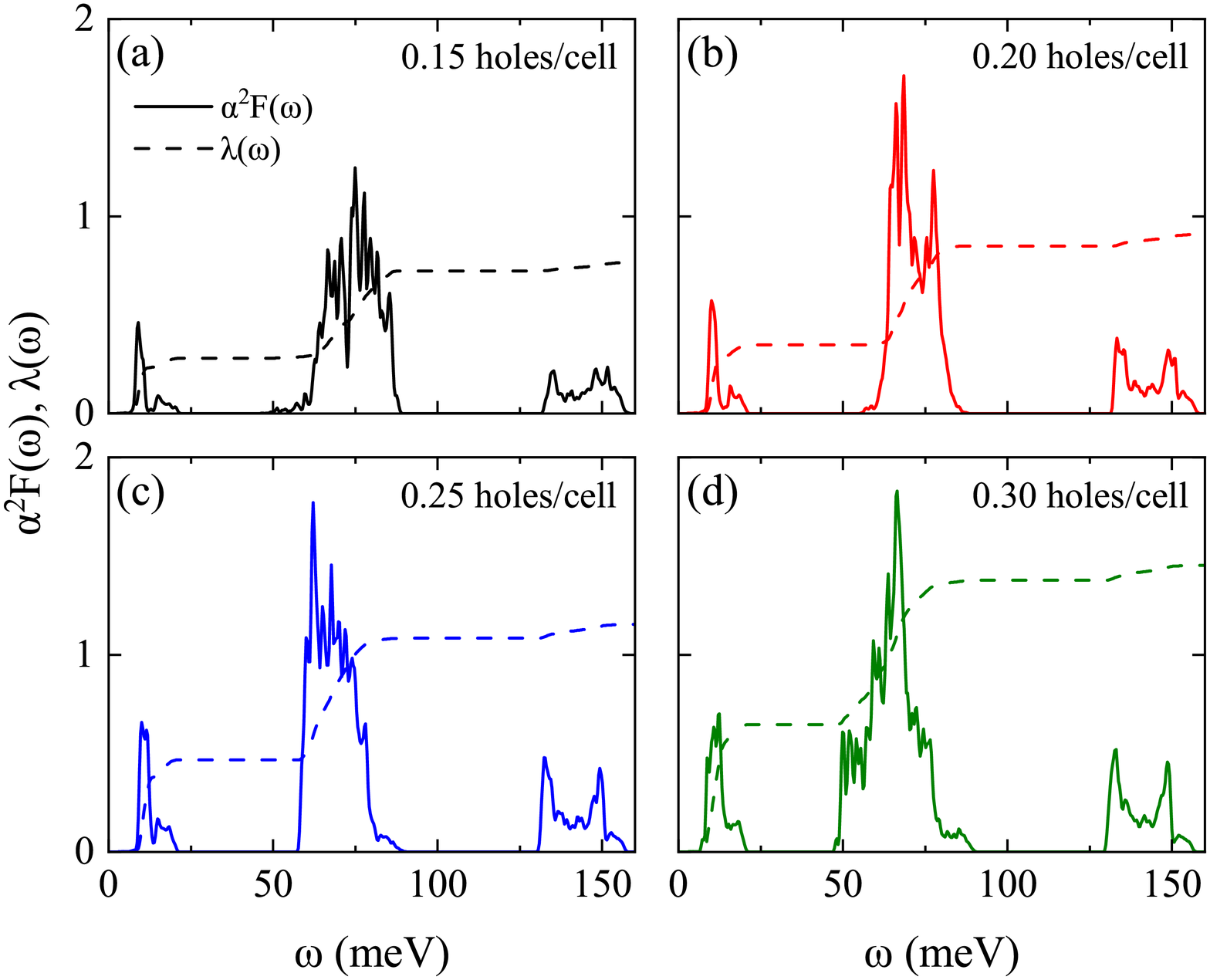}
\includegraphics[width=8.6cm]{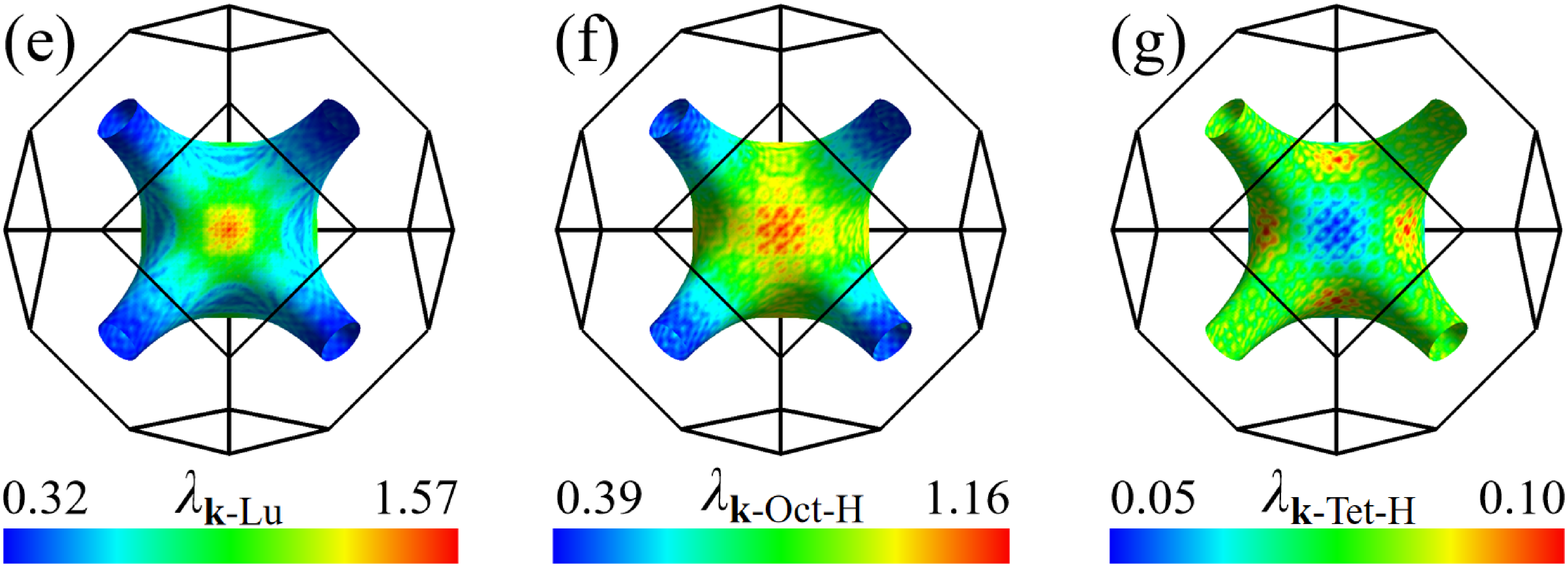}
\caption{(a)-(d) Eliashberg spectral function $\alpha^2F(\omega)$ and accumulated EPC constant $\lambda(\omega)$ for 0.15, 0.20, 0.25 and 0.30 holes/cell doped LuH$_3$, respectively. The accumulated $\lambda(\omega)$ is computed using the fomula $\lambda(\omega)=2 \int_{0}^{\omega}{\frac{\alpha^2F(\omega')}{\omega'}} d\omega'$. The solid and dash lines denote the $\alpha^2F(\omega)$ and $\lambda(\omega)$. (e)-(g) Distributions of EPC strength $\lambda_{\bf{k}}$ originated from three $\alpha^2F(\omega)$ peaks.}
\label{fig:a2f}
\end{figure}

The Eliashberg spectral functions $\alpha^2F(\omega)$ for hole-doped LuH$_3$ are shown in Fig.~\ref{fig:a2f}.
There are three peaks in $\alpha^2F(\omega)$, associated with Lu, Oct-H, and Tet-H phonons, according to the projected phonon DOS. By integrating $\alpha^2F(\omega)$, the EPC constant $\lambda$ can be obtained.
For convenience, we label the strength of EPC that originate from these three peaks as $\lambda_\text{Lu}$, $\lambda_\text{Oct-H}$, and $\lambda_\text{Tet-H}$, respectively. Here, we have $\lambda_\text{Lu}$+$\lambda_\text{Oct-H}$+$\lambda_\text{Tet-H}$=$\lambda$. For 0.15 holes/cell doped LuH$_3$, $\lambda$ is calculated to be 0.77, whereas
$\lambda_\text{Lu}$, $\lambda_\text{Oct-H}$, and $\lambda_\text{Tet-H}$ are 0.28, 0.45, and 0.04, contributing 36.4\%, 58.4\%, and 5.2\% to the total $\lambda$. This means that Tet-H phonons do not effectively participate in EPC.
From 0.20 to 0.30 holes/cell, the EPC constants are found to be 0.91, 1.15, and 1.45, respectively. Especially, $\lambda_\text{Lu}$, $\lambda_\text{Oct-H}$, and $\lambda_\text{Tet-H}$ are equal to 0.64, 0.73, and 0.08 under 0.30 holes/cell doping. This strongly suggests that Lu and Oct-H phonons play an important role in boosting $\lambda$. We also show the {\bf k}-resolved distributions of $\lambda_\text{Lu}$, $\lambda_\text{Oct-H}$, and $\lambda_\text{Tet-H}$ on the Fermi surfaces, to distinguish the electronic states that are strongly coupled with phonons [Fig.~\ref{fig:a2f}(e)-Fig.~\ref{fig:a2f}(g)].
As indicated, Lu and Oct-H phonons have strong interaction with electrons around the face centers and the saddle points, which mix Lu-$5d$ and Oct-H-$1s$ orbitals together,
compared with the orbital-resolved Fermi surfaces shown in Fig.~\ref{fig:band}(c)-Fig.~\ref{fig:band}(d).

\begin{figure}[tbh]
\centering
\includegraphics[width=8.6cm]{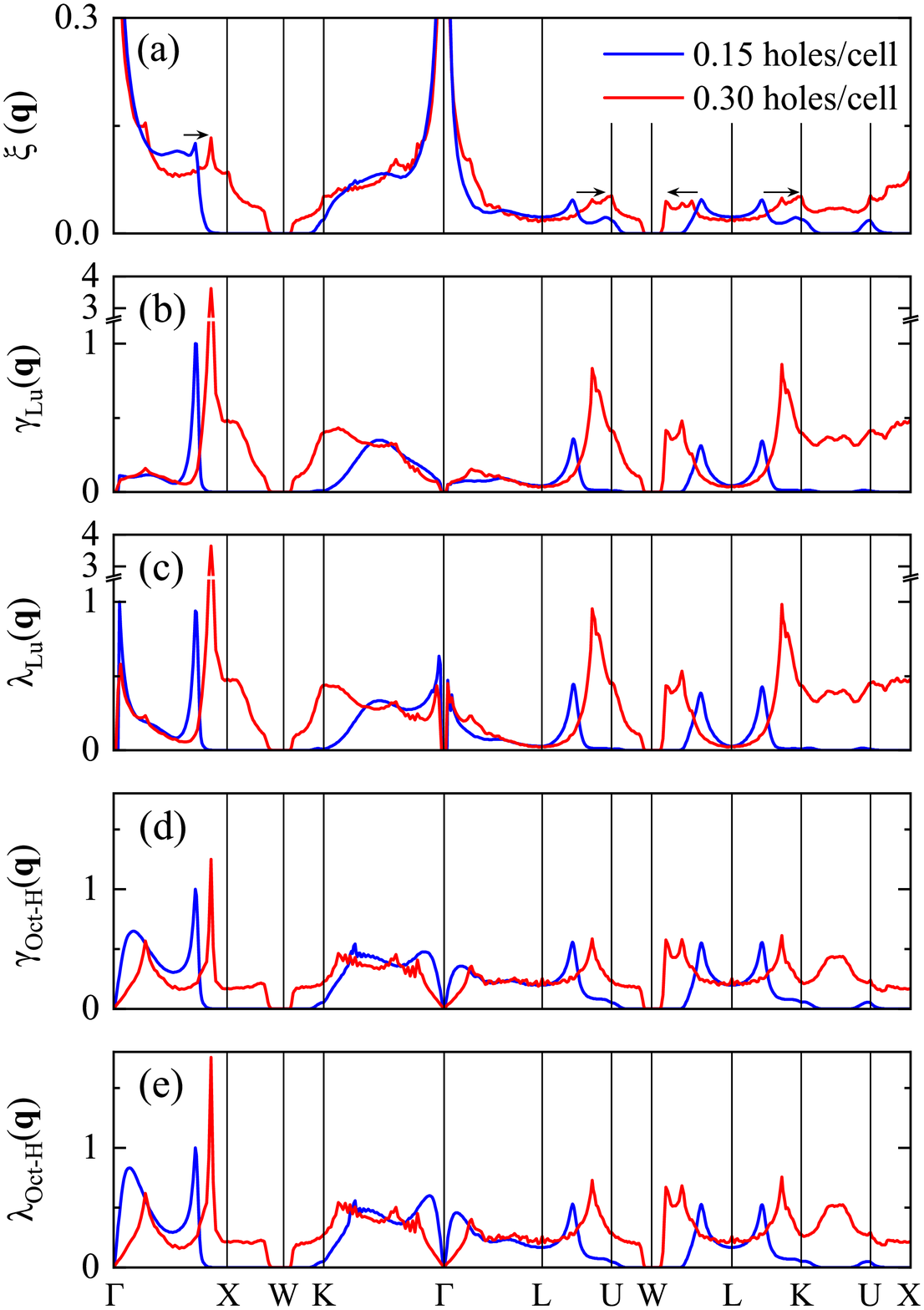}
\caption{Nesting function $\xi({\bf q})$, atom-related $\gamma({\bf q})$ and atom-related $\lambda({\bf q})$ under two doping limits. Here, $\xi({\bf q})$ is normalized by $\xi({\Gamma})$ of 0.15 holes/cell doping. $\gamma({\bf q})$ and $\lambda({\bf q})$ are normalized by their respective maximal value in 0.15 holes/cell doping.}
\label{fig:nesting}
\end{figure}

To figure out the physical reasons that account for the enhanced $\lambda$, we further calculate the Fermi surface nesting function $\xi({\bf q})$, EPC matrix element weighted nesting function $\gamma({\bf q})$,
and {\bf q}-resolved EPC constant $\lambda({\bf q})$. $\xi({\bf q})$, $\gamma({\bf q})$, and $\lambda({\bf q})$ read
\begin{equation}
\xi(\mathbf{q})=\frac{1}{N(0)N_\mathbf{k}}\sum_{nm\mathbf{k}}\delta
(\epsilon _{\mathbf{k}}^{n})\delta (\epsilon _{\mathbf{k+q}}^{m}),
\label{eq:xi}
\end{equation}
\begin{equation}
\gamma(\mathbf{q})=\frac{1}{N(0)N_\mathbf{k}}\sum_{nm\mathbf{k}\nu}|g_{\mathbf{k},\mathbf{q}\nu }^{nm}|^{2}\delta
(\epsilon _{\mathbf{k}}^{n})\delta (\epsilon _{\mathbf{k+q}}^{m}),
\label{eq:gamma}
\end{equation}
and
\begin{equation}
\lambda({\bf q})=\sum_\nu\lambda_{{\bf q}\nu}=\frac{2}{\hbar N(0)N_\mathbf{k}}\sum_{nm\mathbf{k}\nu}\frac{1%
}{\omega _{\mathbf{q}\nu }}|g_{\mathbf{k},\mathbf{q}\nu }^{nm}|^{2}\delta
(\epsilon _{\mathbf{k}}^{n})\delta (\epsilon _{\mathbf{k+q}}^{m}),
\label{eq:lambda}
\end{equation}
respectively.
$\omega _{\mathbf{q}\nu }$ is the vibrational frequency of phonon ${\bf q}\nu$. $m$ and $n$ stand for the indices of energy bands.
$g_{\mathbf{k},\mathbf{q}\nu }^{nm}$ denotes the EPC matrix element. $\epsilon_{\mathbf{k}}^{n}$ and $\epsilon_{\mathbf{k+q}}^{m}$
are the eigenvalues of the electronic states. $N_\mathbf{k}$ represents the total number of {\bf k} points in the fine mesh.

\begin{figure}[tbh]
\centering
\includegraphics[width=8.6cm]{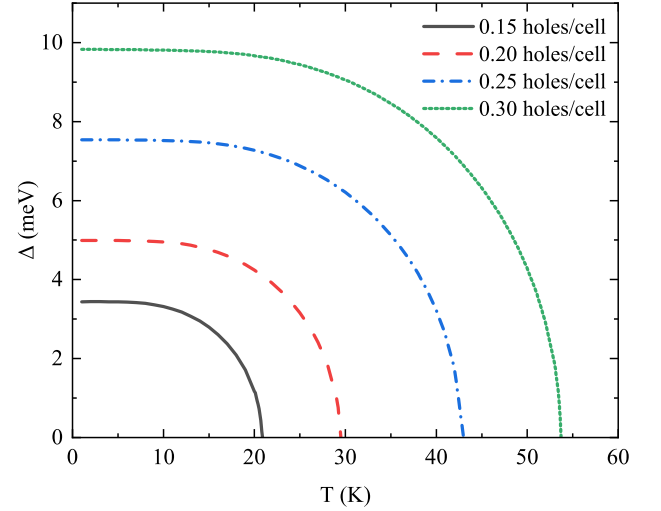}
\caption{Temperature dependence of the isotropic superconducting energy gap $\Delta$ for LuH$_3$ at different hole doping levels.}
\label{fig:Tc}
\end{figure}

For these two doping limits, the peak values of nesting function are almost unchanged, but the peak positions are shifted due to enlarged Fermi surfaces, as indicated by black arrows [Fig.~\ref{fig:nesting}(a)].
The shifted peaks lead to enhanced $\xi({\bf q})$ in partial region of the Brillouin zone, for instance, around the $X$, $W$, $K$, and $U$ points. At these regions, the $\gamma({\bf q})$ of Lu phonons in 0.30 holes/cell doped LuH$_3$ are further amplified [Fig.~\ref{fig:nesting}(b)]. In particular, the highest peak of $\gamma({\bf q})$ under 0.30 holes/cell doping is three times that of 0.15 holes/cell doped case.
The situation in $\lambda({\bf q})$ associated with Lu phonons resembles $\gamma({\bf q})$ [Fig.~\ref{fig:nesting}(c)]. These observations suggest that the increscent EPC matrix elements, rather than the phonon softening, play an important role in strengthening the coupling between Lu phonons and electrons.
For the Oct-H phonons, there is no significant variation in $\gamma({\bf q})$ with respect to $\xi({\bf q})$, except for nearby the $\Gamma$ point [Fig.~\ref{fig:nesting}(d)].
This means that the EPC matrix elements of Oct-H phonons are not sensitive to the doping concentration.
Compared with $\gamma({\bf q})$, peaks in $\lambda({\bf q})$ are raised [Fig.~\ref{fig:nesting}(e)], indicating that the softening of Oct-H phonons is critical to promote their contribution to EPC.

Both the enlarged EPC matrix elements of Lu phonons and the softening of Oct-H phonons can give rise to the red shift of the spectral weight of $\alpha^2F(\omega)$.
Simultaneously, the logarithmic average frequencies $\omega_{\text{log}}$ are reduced from 36.90 to 36.44, 34.56, and 30.70 meV, accompanied by the increase of doping concentration.
By solving the isotropic Eliashberg equations, we can determine the $T_c$ of hole-doped LuH$_3$ [Fig.~\ref{fig:Tc}].
Interestingly, hole doping not only stabilizes LuH$_3$, but also drives it into superconducting states.
The superconducting transition temperatures are calculated to be 20.9, 29.5, 43.0, and 53.7 K, with the isotropic superconducting energy gaps $\Delta$ being 3.44, 4.99, 7.54, and 9.83 meV at 1 K, respectively.
The positive dependence of $T_c$ on the hole concentration can be interpreted in terms of enhanced EPC constant $\lambda$.

\section{CONCLUSION}

In summary, we have performed first-principles density functional calculations to investigate the influence of doping on dynamical stability of LuH$_3$ at ambient pressure.
We manipulate the total charge to simulate electron/hole doping. As uncovered by our calculations, doping holes is critical to realize ambient-pressure stability.
Utilizing Wannier interpolation method, the EPC and phonon-mediated superconductivity in hole-doped LuH$_3$ are computed. The phonons of Lu and H atoms at the octahedral center provide glue to pair electrons. Specifically, the strongest EPC can be achieved under 0.30 holes/cell doping, leading to superconductivity with $T_c$ being 53.7 K. To the best of our knowledge, this is the first theoretical report about ambient-pressure superconductivity in doped LuH$_3$.

\begin{acknowledgments}

This work was supported by the National Natural Science Foundation of China (Grant Nos. 11934020 and 11974194).
Computational resources were provided by the Physical Laboratory of High Performance Computing at Renmin University of China.

\end{acknowledgments}

\end{document}